**Chapter**

# Blockchain Integrated Federated Learning in Edge-Fog-Cloud Systems for IoT based Healthcare Applications: A Survey

***Shinu M. Rajagopal[1], Supriya M.[2], Rajkumar Buyya[3]***

*[1,2]Department of Computer Science and Engineering, Amrita School of Computing, Bengaluru, Amrita Vishwa Vidyapeetham, India*
*[3]CLOUDS Lab, School of Computing and Information Systems, The University of Melbourne, Australia*
**Corresponding author. E-mail: mr_shinu@blr.amrita.edu*

**ABSTRACT**

Modern Internet of Thing (IoT) applications generate enormous amount of data. To prototype the incoming data from such applications, data-driven machine learning has emerged as a viable method which can help to develop precise and reliable statistical models. Existing data is not effectively utilized by machine learning as it is stored in data silos and technology businesses are now required to treat user data carefully in accordance with user-privacy legislation in many regions of the world. It is self-evident that the samples in the typical machine learning centralized server paradigm have different probability distributions of data supplied. As a result, the common model fails to personalize. Without sufficient data, machine learning is unlikely to reach its full potential. Hence the new distributed paradigm, federated learning has gained popularity that supports collaborative learning while preserving privacy. IoT applications with cryptography characteristics will be capable of storing and transmitting data securely over networks by maintaining data consistency. Therefore, federated learning and blockchain integration are particularly beneficial for IoT applications that manage sensitive data such as healthcare. Although several studies have focused on various applications of blockchain technology and federated learning, a thorough examination of these technologies in edge-fog-cloud based IoT computing systems and healthcare applications is yet to be performed. The basic architecture, structure, types, functions, and characteristics of federated learning and blockchain are addressed at the outset of this survey article. This article also considers the wide range of applications to which federated learning and blockchain has been implemented, including edge, fog, cloud and IoT computing paradigms. Lastly, it evaluates the different implementations of federated learning in healthcare applications.

## 1 INTRODUCTION

A centralized system is undesirable for the business and the environment in the long run and hence it can lead to monopolization by a few big companies. It will eventually limit the participation of smaller enterprises. The next level of artificial intelligence is built on the concept of data privacy as its foundation and the users' concern about the privacy of available data such as personal identifiable details, payment data, protected health information, confidential data. Two more challenges that occur with cloud/centralized based techniques are latency and data transfer cost, as the data transmit through a network, in and out of the cloud. This could be solved by federated learning which is a machine learning technique which requires training an algorithm on a few distributed servers or edge devices that retain local data samples without communicating them.

By removing the need to store data in the cloud, mobile devices can jointly develop a prediction model using federated learning while retaining all the training data on the device. Federated learning (FL) enables smart models, reduced latency, and reduced battery usage while protecting one's security and privacy. Training occurs only when the device is powered up, connected to a Wi-Fi network and idle, ensuring that the phone's performance is unaffected. It works with the concept of training the statistical models through distant devices or data centers, like hospitals or smart phones, while maintaining local data. Defense, telecommunications, pharmaceutics and IoT are just a couple of applications where it can be used. The federated learning involves from tens to millions of distant devices combined into a single, global statistical model. Smart phones, wearable technologies and self-driving vehicles are few examples of the current distributed networks that generate massive amounts of data daily. Because of the rising computational capacity of these devices, as well as issues about transferring confidential data, local data storage and pushing of network computing to the edge is becoming more desirable. The term device, used here, refer to nodes, clients, sensors, and organizations that are part of the communication network. Users may be unwilling to reveal their information to protect their personal data or conserve their phone's limited bandwidth/battery capacity. This model is with the limitation that device-generated data is stored locally and processed, and only intermediate modifications are transferred to a central server on a periodic basis. Only a limited subset of devices in federated networks typically engage in each round of training (T. Li et al., 2020).

Federated learning makes use of an aggregator, that is a centralized server which combines locally trained models including model updates. In such systems, the presence of an aggregator raises various significant challenges. The aggregator may become a possible single point of failure. If it fails, the entire training process collapses. The parties will be unable to cooperate on an unreliable aggregator. To address these problems, there have been several decentralized federated learning architectures established. One of the most powerful decentralized architectures for addressing the issue of a secure central aggregator is blockchain. With advanced characteristics like tamper-proofing, confidentiality, and traceability, blockchain has earned a lot of interest for boosting security in areas like IoT, which can be combined with federated learning technology to provide low power consumption, low latency, and privacy (Nguyen et. al, 2021).

A blockchain is a series of blocks that are linked together to store all committed transactions on a public ledger. These blocks contain the entire list of transactional data. The initial block is called the genesis block as it has no parent. Each other block has a relation to the block before it, which is simply a hash value of the previous block called parent block. The transaction size and block size together identify the most transactions that can be accommodated in a block (Z. Zheng et al., 2018). Without the need for a regulatory body, blockchain allows a number of people to reach an agreement on a specific action and record the agreement. None of the approved and registered activities can alter without the participation of the other participating bodies (Al-Jaroodi & Mohamed, 2019). In a decentralized and distributed network, a blockchain is a collection of information-storage blocks with digital signatures. (i.e.) a multi-field infrastructure technology (Lin & Liao, 2017). This is supported by a number of fundamental technologies, including distributed consensus methods, cryptographic hashes, and digital signatures. decentralization, immutability, accountability, and auditability are all features of blockchain that make transactions highly safe and tamper-proof (Monrat et al., 2019). Because of the public trust that it encourages, blockchain enables the removal of intermediary third parties (Berdik et al., 2021). Privacy, protection, computation, power consumption, network traffic, storage, scalability, and technological abuse are all challenges in blockchain.

The decentralized digital currency, Bitcoin, a version of blockchain's distributed ledger, has a block size restricted to one megabyte with the total block formation duration of ten minutes, limiting the network's throughput (Tang et al., 2020). Bitcoin's transaction processing speed ranges from 3.3 to 7 transactions per second. Problems arise over a variety of issues, including the transaction delays, blockchain congestion and increased transaction charges. Thus, relying on the blockchain platform to build a business model may not be a viable option for the government or private sector. The most popular consensus mechanism currently utilized in blockchain technology is called Proof of Work (PoW). Even so, one of PoW's big drawbacks is the waste of computational power. Ethereum is working on a hybrid consensus framework that combines PoW and Proof of Stake (PoS) to solve this challenge (X. Li et al., 2020). Numerous studies have focused on applying blockchain in numerous application areas, but there isn't a comprehensive study on blockchain technology's use in the edge/ fog/ cloud/ IoT healthcare applications.

The contributions of this study are as follows:

- Overview and fundamentals of federated learning and blockchain concepts
- Identified and analyzed the survey papers of federated learning and blockchain
- Analyzed the research papers of blockchain and federated learning in different application domains using edge/fog/cloud/IoT and identified the next directions for blockchain research based federated medical IoT applications

The rest of this survey paper is organized as follows. Section 2 introduces blockchain and federated learning architecture. Section 3 discuss the significance of federated learning and blockchain integration. Blockchain and federated learning survey papers are examined in Section 4. Section 5 present the federated learning and blockchain application areas. Section 6 summarizes the blockchain based federated learning and Section 7 deals with blockchain and federated learning in healthcare applications. Section 8 concludes the paper.

# 2 PRELIMINARY - FEDERATED LEARNING AND BLOCKCHAIN ARCHITECTURE

## 2.1 Federated learning

This section starts with a general overview of federated learning technology, including its structure, types, and how they function.

### 2.1.1 Overview

Federated learning facilitates machine learning algorithms to gain knowledge from a diverse collection of data sets stored in distributed locations. It is a decentralized, privacy preserving strategy that preserves data on devices and uses local machine learning based training to reduce communication overhead. Federated learning allows collective model training without the need to share training data explicitly with all others involved in the network. It requires an aggregator which is a central system that combines locally trained models and updates. This allows various parties to train a model together without exchanging data or allow the aggregator to infer original data thus lowering the high cost of data gathering. Federated learning is based on an iterative mechanism of client server interactions. Each iteration of this procedure entails sending of the participating nodes with the current global model state and training localized models on such nodes to generate a collection of possible model updates, aggregating and processing these local changes and later processing these local updates (Ng et al., 2021). The federated learning system is depicted in Figure 1.

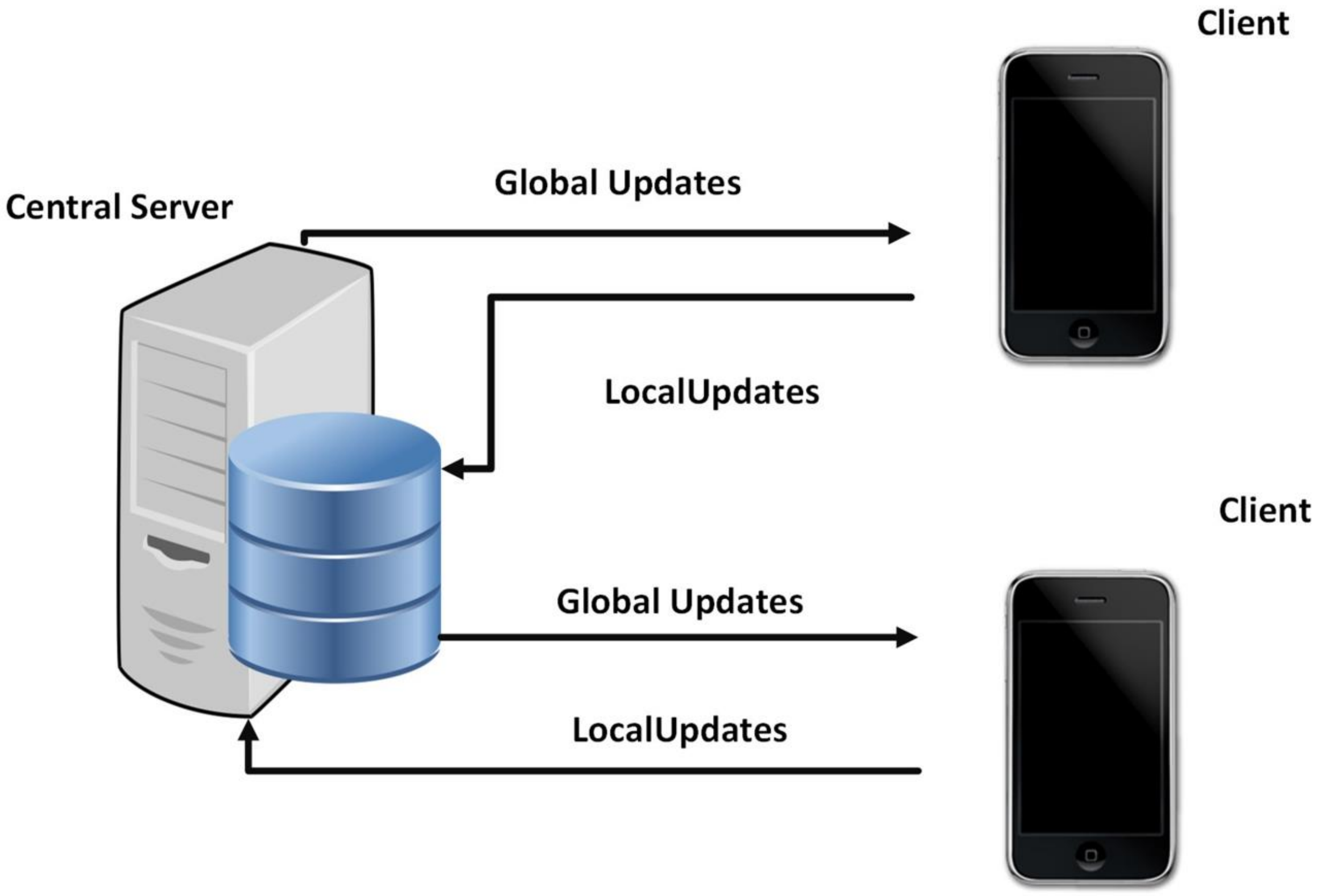

**FIGURE 1 Federated learning system**

### 2.1.2 Structure

To train the machine learning model, there are three major components: clients, servers, and the communication-computation system.

- Clients: Organizations or smart devices that submit local model updates to a central server for aggregation to make practical real-time predictions.
- Servers: It is the aggregator who incorporates local updates to create a global model iteratively to improve the predictions.
- Communication-computation system: On client and server, computation for model training and communication for exchanging model parameters take place. Hence the efficient communication-computation framework is needed for the federated learning.

### 2.1.3 Types

Types of federated learning classes based on training data dispersed throughout the sample and feature spaces are vertical federated learning, horizontal federated learning, and federated transfer learning.

- Vertical federated learning: To train a global model, it incorporates datasets from different feature spaces.

- Horizontal federated learning: Across all platforms, it uses datasets of the same featurespace.
- Federated transfer learning: It is learning with a pretrained model for solving a different problem that was trained on a similar dataset.

### 2.1.4 Functions

- The server identifies a subset of clients to collaborate with. Although the usual conditions for device selection are that the device is charged, idle, and connected to an unmetered network, selected few participants receive the server's most recent model weights and by using them, set up the local machine learning model. The global model is trained and optimized using the local training data of each chosen customer. To compute the update, the client uses stochastic gradient descent technique.
- To improve model performance updates and to minimize communication costs, several gradient descent steps over numerous epochs are processed in one round.
- The clients submit the optimized values to the server, once training is over. Because of poor connection, limited computing resources, large amounts of training data, and other factors, some clients can drop out during the training or parameter transmission cycles. The current active round will be suspended if the number of clients who report in time is insufficient.
- After weighing the updates depending on the scale of the dataset, the server aggregates them and a new shared model is developed, which will be improved in consecutive iterations.

### 2.1.5 Features

- Security and Privacy Benefits: Personal data is stored locally on one’s personal server and hence no concern with encryption or privacy.
- Smarter realistic predictions: Since the data sets are available, generating the model does not require a centralized server, more realistic predictions can be made on our computer.
- Lower latency: On the user’s computer, the modified model can be used to make predictions.

### 2.1.6 Challenges

- Data that is not independent or non-identically distributed: Each client creates its own dataset based on individual actions and system use.
- Massive distributed unbalanced data: The large amount of varying training data is generated by the system’s distributed heterogeneous clients.
- Intermittent device Connection: The level of network access varies considerably from one client to the next which causes intermittent behavior sometimes.
- Device Memory and processing constraints: The end devices used in the learning process has

processing power and memory limitations.

- Security threats: Because of the anonymity of the clients, an intruder may pose as a regularuser and be chosen to participate in the learning process.

## 2.2 Blockchain

This section starts with a general overview of blockchain technology, including its structure, types of blockchains, and how they function.

### 2.2.1 Overview

With the publication of the Bitcoin whitepaper in 2008, the blockchain idea became widely known (Nakamoto, 2008). Blockchain is a chain of connected data blocks, each one dependent on the one before it, creating a continuous data structure. A blockchain is a permanent and verifiable ledger of transactions. A block contains a hash of the previous block which influences the chain. The blockchain's size is decreased by using the Merkle tree structure by representing the transactions as leaf nodes, each of which is separately hashed. Until the root is reached, each group of child hashes is concatenated and hashed once more up the tree (Gao et al., 2018). Transactions are broadcast across the network once they are generated. The timestamp confirms that the transactions occurred at the time the block was created. The nonce is a onetime-use number that must be calculated therefore the current block's hash value satisfies some arbitrary conditions, such as starting with a predetermined number of zeros to impose the complexity of calculating a block's hash. When blocks are added to the blockchain, the nodes participating in transactions and block creation must be able to confirm their validity. Simplified Payment Verification (SPV) is a payment authentication technology that uses block header messages rather than complete blockchain information to verify payments. This has the potential to significantly reduce user storage in blockchain payment verification, as well as relieve user burden as transaction volumes grow massively (Z. Zheng et al., 2018).

### 2.2.2 Structure

The header block includes the following (Berdik et al., 2021):

- Block version specifies which set of block validation rules should be used.
- A 256-bit hash value called the "parent block hash" connects to the block before it.
- The hash value computed for each transaction in the block is known as the "Merkle tree root hash."
- Timestamp is the current timestamp represented by n Bits is the current hashing objective in a concise form.
- The 4-byte field known as Nonce begins with 0 and increases with each hash computation.

The block body includes:

- Transactions counter and transaction.

Structure of Blockchain is depicted in Figure 2.

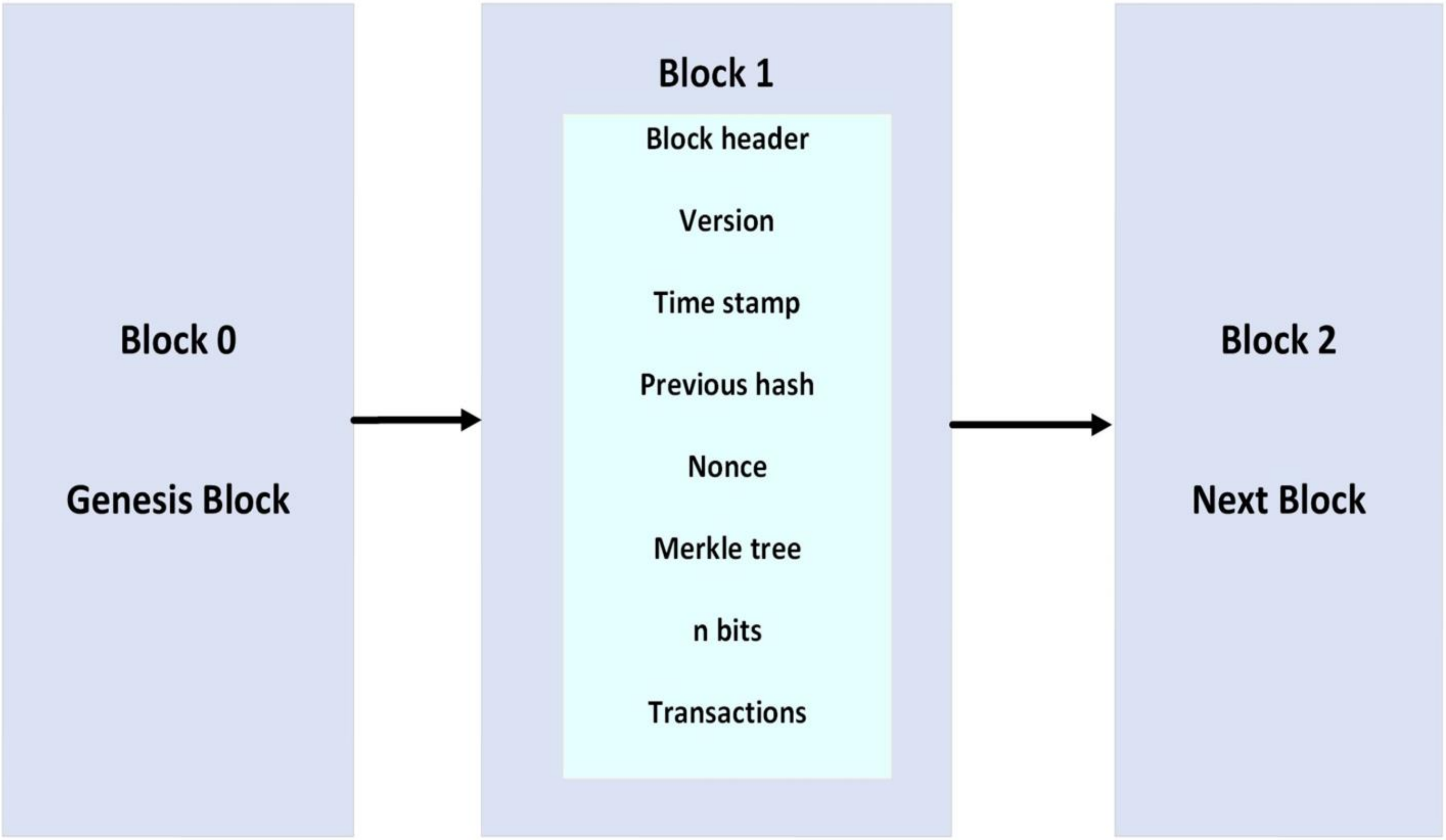

**FIGURE 2 Structure of Blockchain**

### 2.2.3 Types

Blockchain technology can be categorized as follows:

- Private blockchain: This does not allow any node to participate. It has a rigid control management system in place for data access. Only approved nodes are permitted to participate in the private blockchain.
- Public blockchain: All can check and validate the transaction, and they can also engage in the consensus process, as can be seen in public blockchains of Bitcoin and Ethereum.

- Consortium blockchains: The node with authority can be chosen ahead of time, that the blockchain's data can be either public or private, and that partially decentralized consortium blockchains like Hyperledger and R3CEV.

Types of blockchain is described in Figure 3.

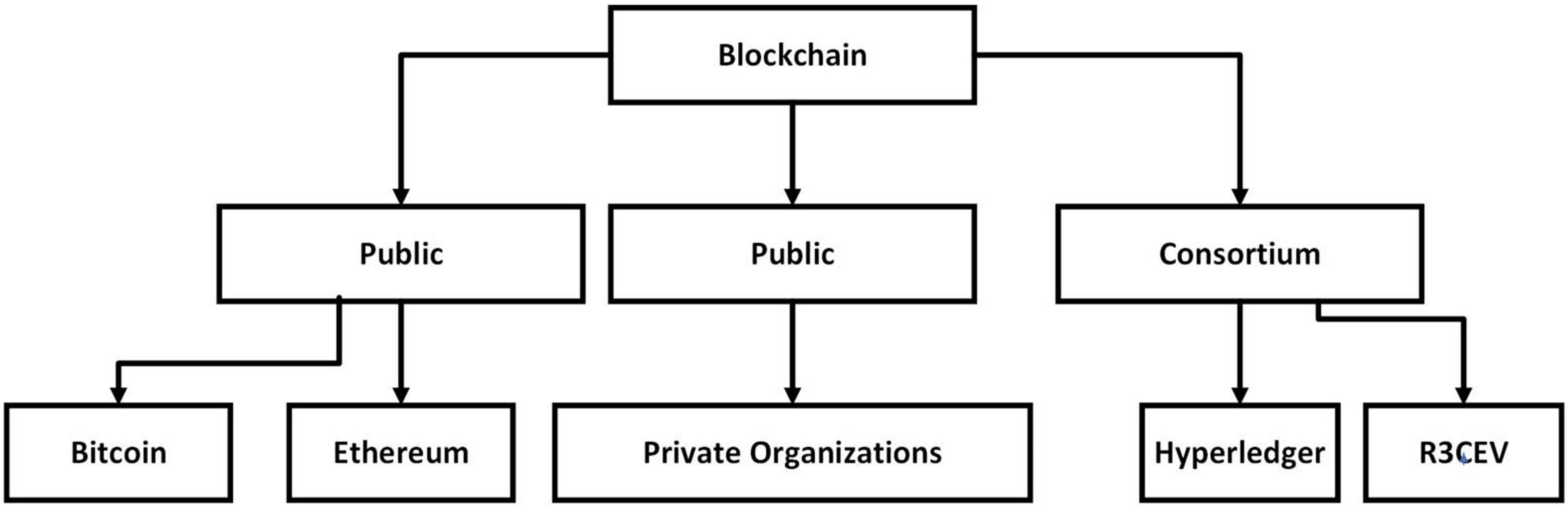

**FIGURE 3 Types of blockchain**

### 2.2.4 Functions

The main working processes of blockchain are as follows:

- The transmitting node reports new transactions and transmit them to the rest of the network.
- Mining operations will be carried out by all mining nodes, and the winner of the puzzle will update the blockchain with a new block and is eligible to get reward.
- After the consensus algorithm has been run, the block will be added to the chain, which every node in the network will accept, allowing the chain to grow.

Functioning of blockchain is presented in Figure 4.

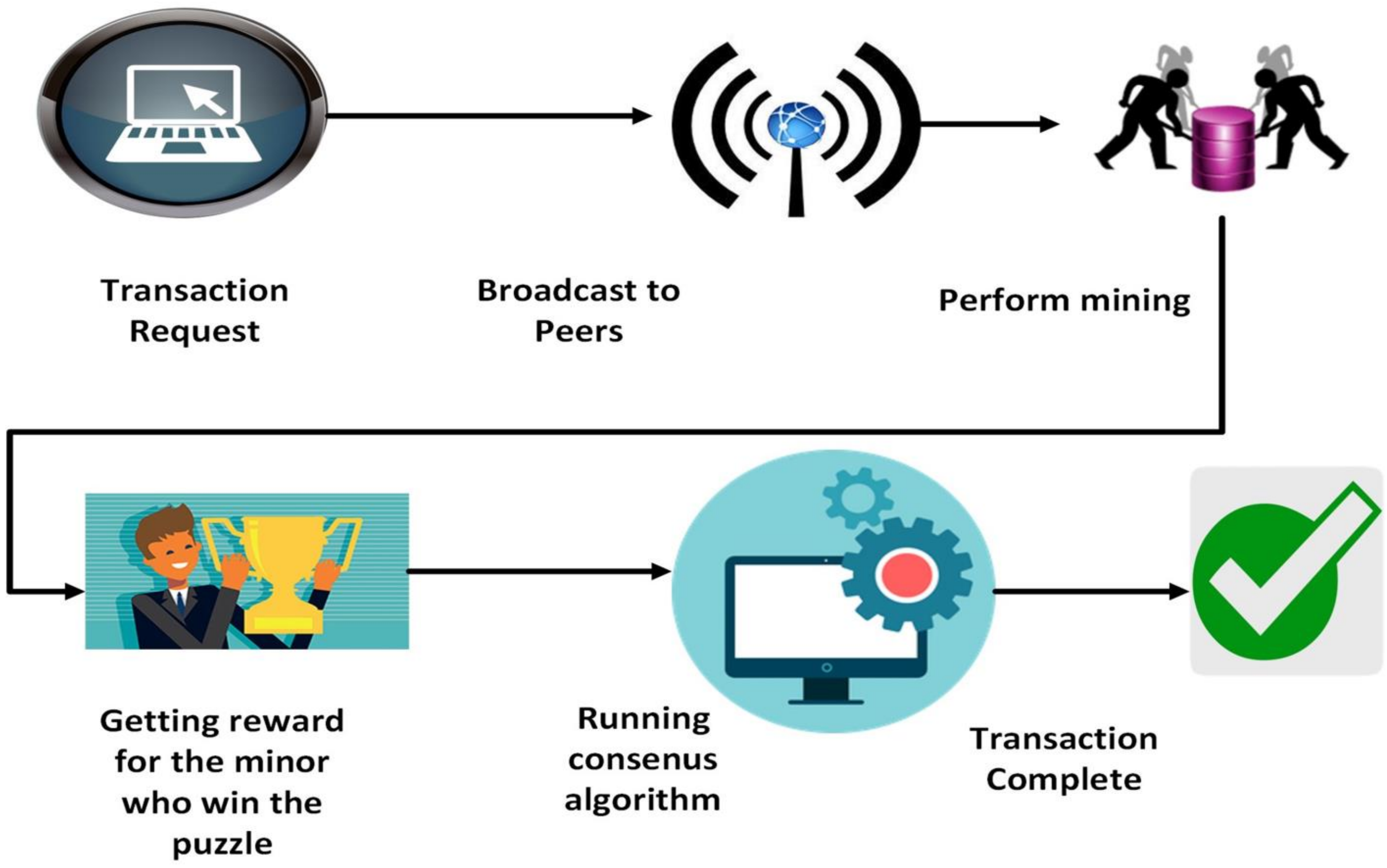

**FIGURE 4 Functioning of blockchain**

### 2.2.5 Features

- System transparency: An application or organization may use blockchain to create a decentralized network that eliminates the need for a central system, which is the system transparency improvement.
- Improved Security: Other nodes in the network would be aware of any malicious attacker trying to change the transaction. Transaction that has been published in the block cannot be modified in any way.
- Cost reduction: There is no need to pay for third-party vendor costs because blockchain is a decentralized framework.
- Automation: This is programmable, and when the trigger's conditions are met, it could automatically execute a series of actions such as payments.
- Data protection: In a public ledger, the chain records the entire non-reversible log of transactions while new transactions are added in immutable manner and old transactions are kept in the ledger permanently.

  Blockchain features description is listed in Figure 5.

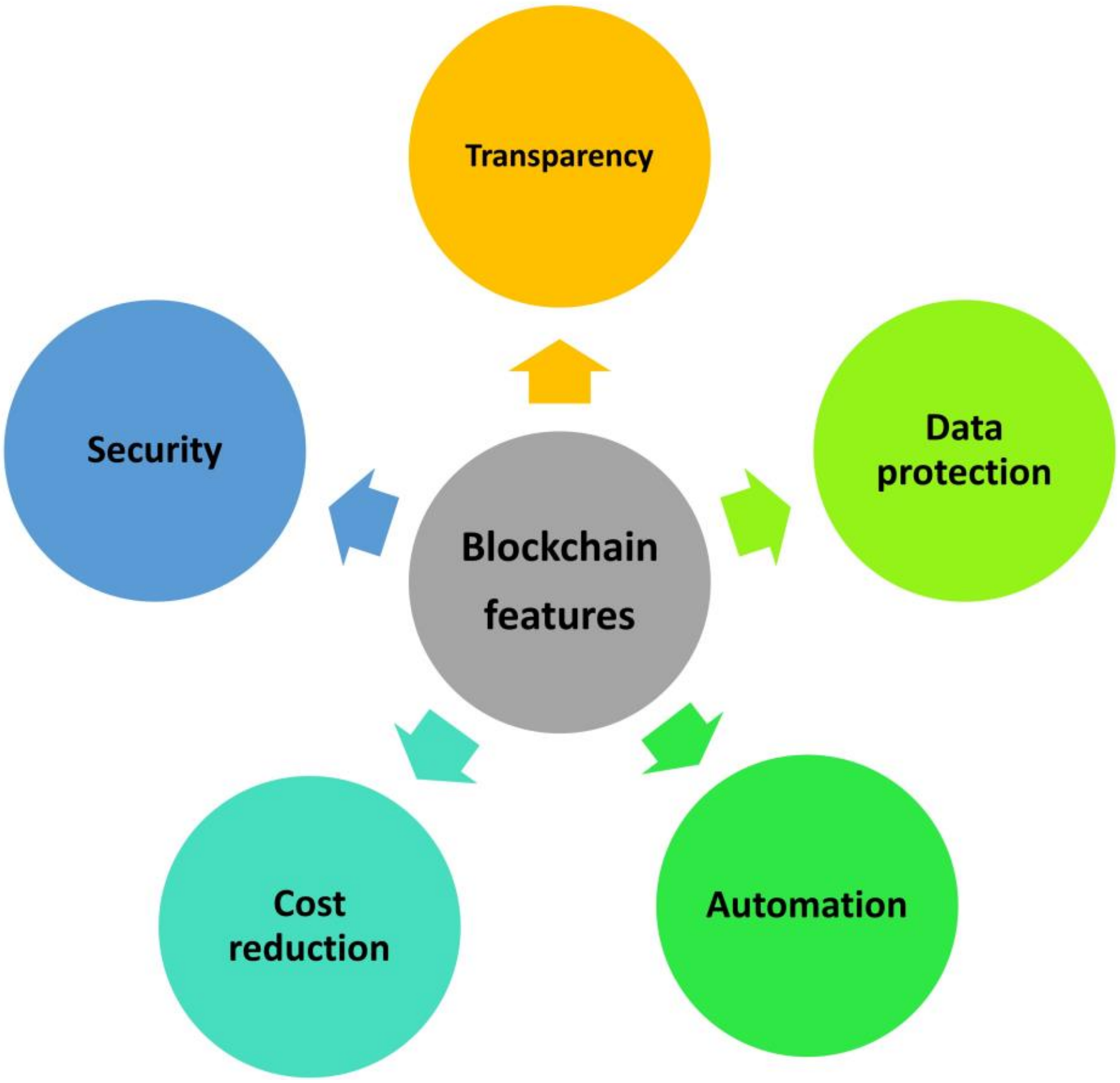

**FIGURE 5 Blockchain features**

### 2.2.6 Challenges

- Scalability: The capacity of the blockchain platform to accommodate growing transaction loads as well as the size of the network's nodes is difficult to achieve.
- Energy consumption: Mining for consensus algorithms would force machine to solve complex equations, which will consume more energy.
- Memory requirements: The blockchain platform's memory capacity to handle growing transactions is challenging.

## 3. BENEFITS OF INTEGRATING FEDERATED LEARNING WITH BLOCKCHAIN

To meet the demands of the modern world, IoT has the capacity to handle huge amounts of data transmission, storage, and real-time processing. To deal with IoT applications, artificial intelligence, fuzzy logic and machine learning approaches are used. However, these methods require a lot of computing power, which is not practical in a completely distributed system. A centralized design is not adequate for the accurate decision making required by many real time IoT applications within a given time frame. FL's distributed machine learning technique makes it the best option to solve such issues. Mobile devices gather data, which they then utilize to create local models and train them. Then, local model data are given to the aggregator, which creates a global model by averaging the local models. Each global model is trained on mobile devices until the required result is obtained. But FL models have few issues related to the security of data. An attacker can access sensitive data on an end device through model updates which are already on the primary server. Additionally, the malicious end device has the ability to alter the current local model. Additionally, to deliver an inaccurate model update to connected end devices during the execution, the attackers can execute an injection attack against the central server. Due to all the factors mentioned above, FL security is crucial. The combination of FL and blockchain can address the security issues with an existing system, which is especially appropriate for applications in the delicate field of healthcare. When opposed to centralized networks, the decentralized approach will maximize the utilization of computing resources and offer balanced workloads. Additionally, the blockchain guarantees data consistency by utilizing cryptographic properties. IoT apps with cryptography capabilities will be able to store and transmit data securely over networks. Blockchain saves details of medical data, and each patient's identity is assured which ensures the data authentication and authorization. Blockchain performs transaction validation and saves locally trained data of various transactions for medical data, maintaining a secure data transfer. The most difficult problem facing crucial infrastructures like the healthcare system is data breaches. Data encryption methods are provided by blockchain to decrease the risk of data breaches. Providing an FL solution based on the blockchain permits the communication of medical data using encrypted identities deprived of the central server. After adding the transaction to the block, the transaction is dispersed equally among all network nodes. The issues of security and privacy of IoT applications can be solved by FL and blockchain integration (Nguyen et. al, 2021), (Aich et. al, 2022), (Drungilas et al., 2021).

Federated learning for edge computing based medical application is depicted in Figure. 6. This figure helps us to understand the application of federated learning concepts in medical sector. Blockchain based federated learning is presented in Figure 7.

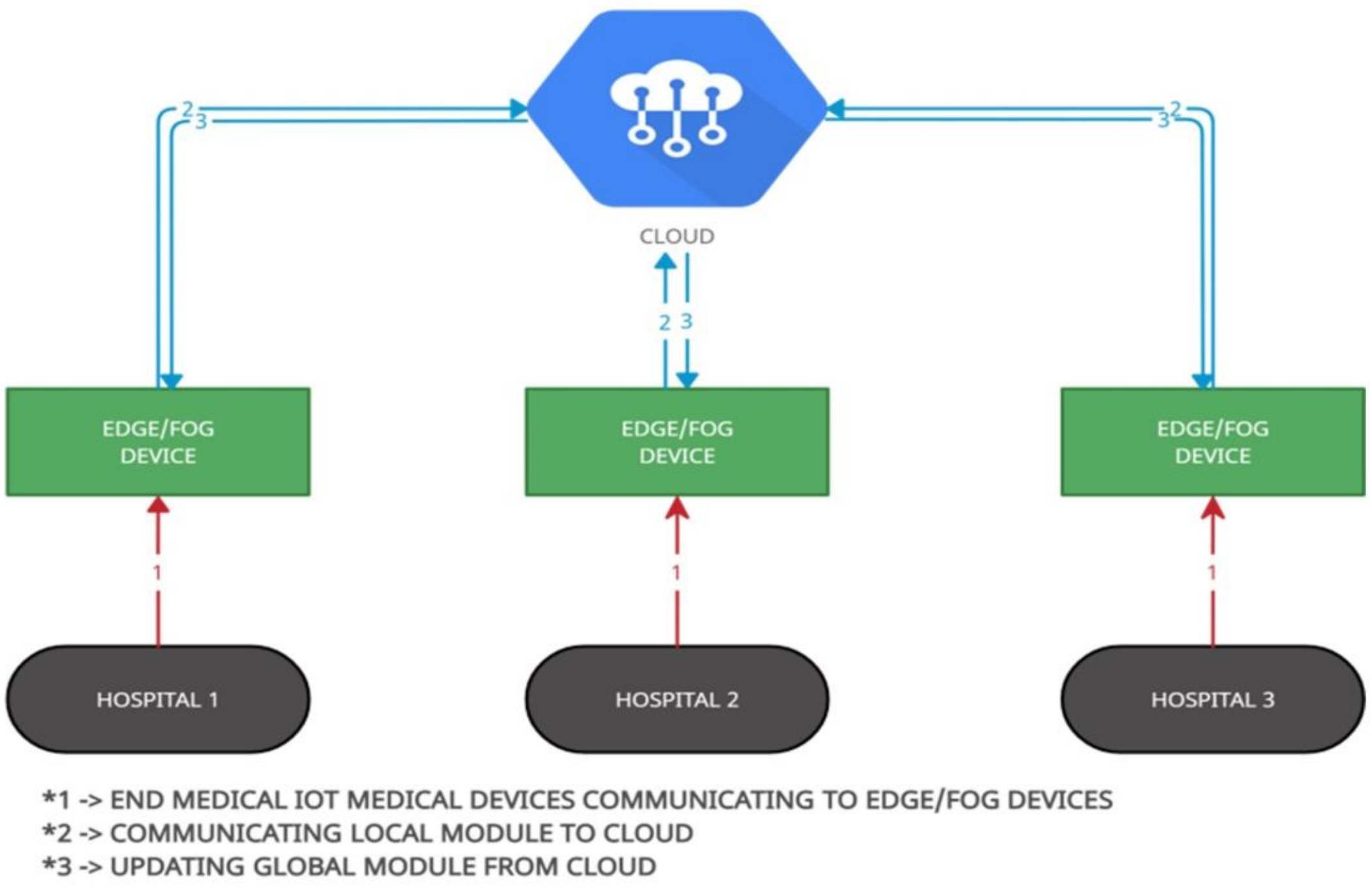

**FIGURE 6** Federated learning for edge computing based medical application

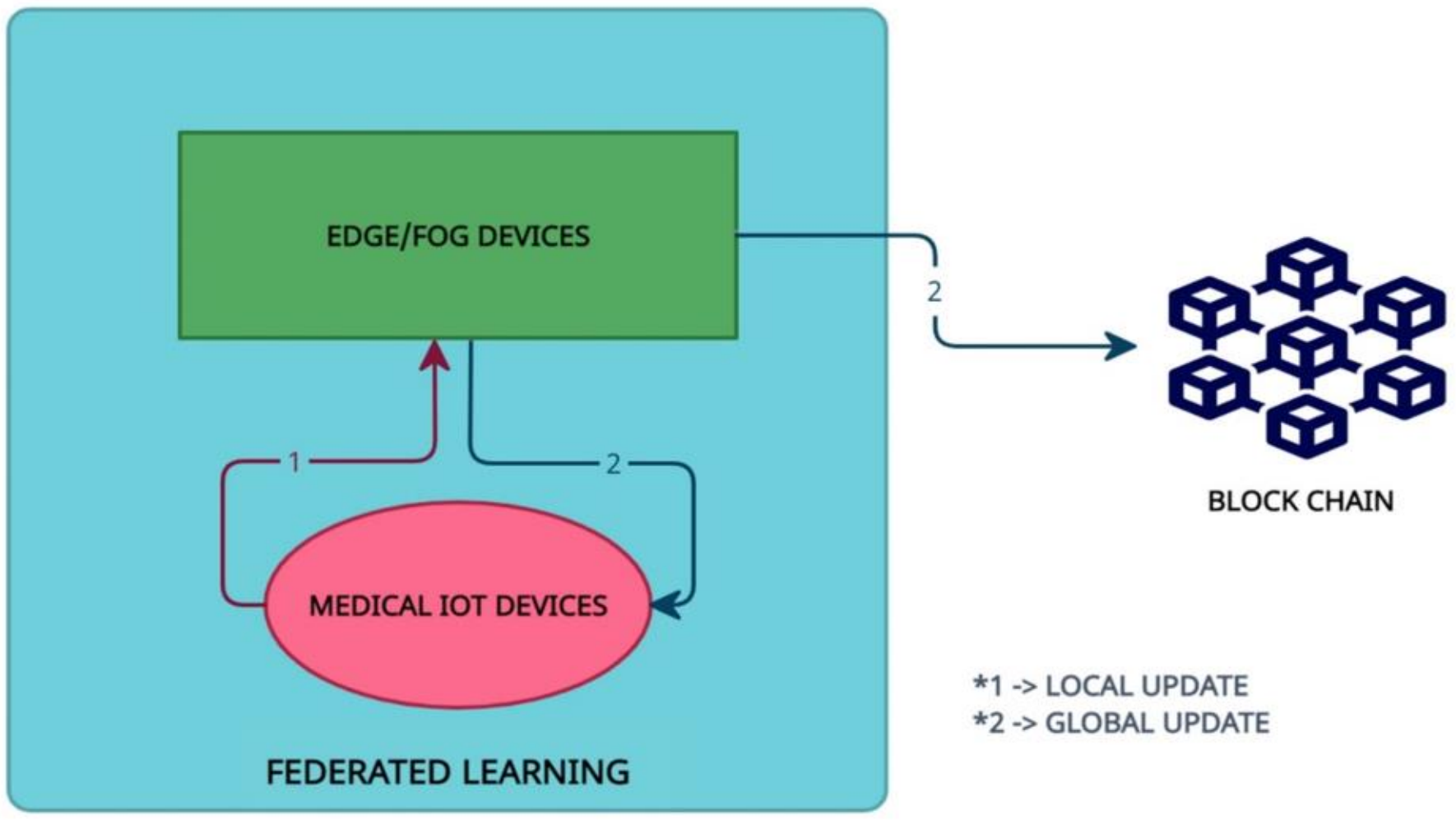

**FIGURE 7 Blockchain based federated learning**

## 4. RELATED WORKS

This section elaborates the existing literature of survey papers in blockchain and federated learning.

A more thorough overview of the most important FL hardware, software, platforms, and protocols help researchers learn about FL quickly. Practical examples of FL applications and use cases to show how different FL architectures can be applied for a variety of scenarios, helping the audience better understand how FL can be used for healthcare and other areas is presented by Aledhari et. al (Aledhari et al.,2020). Liu et. al discusses frequently used federated systems based on the functional architecture and explains federated learning systems from four perspectives: parallelism, aggregation techniques, data transfer, and the security (Liu et al., 2022). According to Abdulrahman et. al, the research directions are divided into four categories: system model and design, application areas, privacy and security, and resource management in the fields of Gboard, smart healthcare, IoT, edge computing, networking, robotics, grid-world, models, recommender systems, cybersecurity, online retailers, wireless communications, and electric vehicles (Abdulrahman et al.,2020). Federated learning research exploration from five perspectives presented by Zhang et. al, addresses data partitioning, privacy mechanisms, machine learning models, communication architectures, and systems heterogeneity. In order to examine the existing practical use of federated learning, the current difficulties and potential research areas have to be addressed (Zhang et al., 2021) and one needs to focus on communication costs, statistical heterogeneity, system heterogeneity, privacy concerns, and other risks (Zhang et al., 2022). Analysis of system design issues for large-scale FL implementation considers the difficulties with resource allocation, communication costs, privacy, and security concerns (Lim et al., 2020).

In addition to the present research on FL, Rahman et. al surveys a new categorization of federated learning concepts and study topics (AbdulRahman et. al, 2020). The literature also presents a thorough examination of FL's increasing applications in IoT networks, current developments in FL, IoT integration and FL's potential for supporting IoT services like data offloading, IoT data sharing, localization, caching, mobile crowdsensing, attack detection, privacy and security, as well as its use in major IoT applications like, unmanned aerial vehicles, smart transportation, smart healthcare, smart cities, and smart industries (Nguyen et al., 2021). In addition, to leverage blockchain to improve FL performance, various FL incentive system implementations, and blockchain based federated learning (BCFL) industrial application scenarios, has been presented (D. Li et. al, 2022).

Blockchain technologies divides into layers such as data, network, consensus,

and application, with different blockchain applications being classified according to their domains. A survey conducts a thorough investigation into blockchain consensus strategies, network architecture, and applications (M. Wu et. al, 2019). The advantages of adoption of blockchain technology in hospital records management, verifiable end to end electronic voting, identity management systems, decentralized notary, supply chain management and access control systems, is presented by Maesa et. al (Maesa & Mori, 2020). A systematic organizational vision of blockchain networks focusing on the characteristics of blockchain networks', use of decentralized consensus and analysis of distributed consensus system design and the reward mechanism perspectives has also been discussed (Wang et. al, 2019). The main objective of Zhou et. al is to categorize and address all existing blockchain scaling solutions. It analyzes various methods and suggest few possible approaches to solve the blockchain scalability problem (Zhou et al., 2020).

The usage of blockchain as a service to safeguard and manage current information systems is thoroughly examined by Berdik et. Al. The survey offers comprehensive details on numerous blockchain research and implementations suggested by the researchers, as well as their individual effects on blockchain and its use in other application scenarios (Berdik et. al, 2021). Feng et. al detail about how privacy concerns pertain to blockchain. It examines blockchain's privacy risks and discusses current cryptographic security mechanisms including anonymity and transaction privacy protection. It also summarizes common blockchain implementations and examines potential research issues that must be tackled to maintain privacy as blockchain is being used (Feng et al., 2019). By examining well-known blockchain systems like Ethereum, Bitcoin, and Monero, Li et. al investigate common blockchain systems to perform a thorough analysis on the security threats, attacks to blockchain, vulnerabilities exploited in attacks, and the causes or potential consequences of each risk or vulnerability (Li et. al, 2020). The blockchain classification, prominent blockchain consensus algorithms, blockchain implementations, technological problems, current developments in overcoming those challenges, and future perspectives in blockchain technology are presented (Zheng et. al, 2018). Blockchain security concerns are divided into two categories, algorithm-based security and hashing operations-based security. Threats at application blockchain protection have been examined, including network design vulnerabilities and privacy concerns. Potential blockchain vulnerabilities have been identified, including cryptographic operation vulnerabilities, identity vulnerabilities, manipulation-based attacks, quantum vulnerabilities, reputation-based attacks, service vulnerabilities, malware attacks, device vulnerabilities, and threats at application blockchain protection (Dasgupta et al., 2019).

By describing the taxonomy and design of the blockchain, comparing various consensus mechanisms, and analyzing aspects like scalability, anonymity, interoperability, energy consumption, and regulatory concerns, Monrat et. al present a comparative analysis of blockchain trade-offs. This comparison focusses on node identity management, energy conservation, the adversary's power that can be tolerated, consensus methods like POW, POS, Delegated POS considered, practical Byzantine fault tolerance, Tendermint for domains like healthcare, voting, governance, identity management, resources, supply chain, stock exchange, education, digital records, and asset monitoring (Monrat et. al, 2019). According to Gao et. al, the blockchain system is a high-level, all-encompassing definition of the blockchain architecture, which is divided into three layers: network, data, and application. Data structures and algorithms in the data layer include hash and hash pointer, digital signature, Merkle tree, and so on.The network layer contains a distributed agreement consensus mechanism on block validity and allow users to update and distribute the blockchain. The application layer illustrates about using blockchain to leverage the ledger, consensus between nodes, cryptographic elements, and smart contracts (Gao et. al, 2018).

## 5. FEDERATED LEARNING AND BLOCKCHAIN APPLICATION AREAS

This section discusses the related literature of blockchain and federated learning in edge/fog/cloud/IoT applications

For customized federated learning, Wu et. al propose PerFit, a synergistic cloud-edge architecture that addresses system, statistical, and model heterogeneity in IoT applications. It explores at emerging personalized federated learning approaches in a case study of IoT human behavior recognition to show how efficient personalized federated learning can be for smart IoT applications that can reduce the negative effects of heterogeneities in various ways (Q. Wu et al., 2020). The most recent developments in federated learning that have been made possible by IoT applications are powered by federated learning and consider a variety of factors, including sparsification, robustness, quantization, scalability, security, and privacy (Khan et al.,2021). Analysis of FLchain applications in well-known MEC fields such as edge data sharing, edge content caching, and edge crowdsensing including study of difficulties in FLchain design, communication cost, resource allocation, incentive mechanism, security, and privacy protection has been discussed (Nguyen et al., 2021). It is suggested to use a neural-structure aware approach to resource management with federated learning that is module based, where mobile clients are assigned to various subnetworks of the global model depending on the status of their local resources, allowing for elastic and effective resource utilization (Yu et al., 2021). To make deep reinforcement learning based outcomes practical and to lower the transmission costs between the IoT devices and edge nodes, the proposed work suggests the use of numerous deep reinforcement learning agents that can be deployed on several edge nodes to signal the decisions of the IoT devices (Ren et al., 2019). To safeguard the privacy of updated local models with a random distributed update and to eliminate security vulnerabilities by a central curator, the work proposes the addition of local differential privacy into federated learning. By using weighted aggregation and updates verification, it performs the convergence quickly (Lu et al., 2019).

To resolve the issue about the lack of control on the posted ledgers, Zhu et. al propose controllable blockchain data management. The proposed approach's inclusion of a trust authority node allows for the elimination of any suspicious acts. The proposed approach's security and efficiency have also been analyzed (Zhu et al., 2019). The integration of blockchain technology with already-in-use cloud technologies, enabling for cloud datacenter reengineering, and the integration of blockchain technology with present cloud systems for increased performance and security have been discussed in Gai et. al (Gai et al., 2020). Zhao et. al suggests using hierarchical crowd sensing to evaluate the test accuracy of federated learning in systems using blockchains with and without differentiated privacy (Zhao et. al, 2020).

Zhang et. al proposes the BCPay outsourcing services, a blockchain based fair payment system in which the service fee is exchanged directly between the customer and the server, eliminating any third-party while maintaining payment fairness against intruders and outsourcing service providers. This implementation employs an all or nothing checking proof protocol and assess the efficiency of proposed BCPay in terms of transactions and computations involved. In terms of transaction size and cost of computation, BCPay is very efficient (Y. Zhang et al., 2018). ProvChain, a blockchain data provenance architecture that provides a cloud storage application with monitoring of every data in real time access while improving privacy, availability, transparency, and the data accountability as a decentralized and trusted cloud data provenance system has been proposed. It uses a file as a unit of data, hence, all cloud object operations are reviewed and registered using blockchain to capture and track cloud data access events. The computing complexity overhead increases as the file size grows (Liang et al., 2017). A reliable blockchain based decentralized resource management system uses a smart contract embedded reinforcement learning approach to reduce the energy used by the request scheduler. It does not require a scheduler, which adds to the cost of energy and reduces the robustness of datacenters. As compared to other standard algorithms, analytical findings on Google traces of cluster and real-world power prices indicate that the proposed solution can dramatically reduce datacenter costs (C. Xu et al., 2017).

A system that uses blockchain to ensure data integrity during the offloading process of a multimedia workflow, and to manage cloudlets based on blockchain for multimedia workflow is developed using the CloudSim framework. To evaluate the benefits of this proposal, the results are compared to benchmark approaches namely, best fit decreasing, first fit decreasing and hybrid computation offloading algorithms (X. Xu et. al, 2020). The NutBaaS platform offers blockchain services such as network implementation and device management, as well as smart contract analysis and testing, over cloud computing environments. Developers should concentrate on the code and explore to learn to adapt blockchain technology that more suitably fit their commercial circumstances using these services, rather than worrying about maintaining the code

(W. Zheng et. al, 2019). Blockchain technology to propose the first certificate less public authentication scheme for cloud storage against delaying auditors presented in literature has better security assurance. The main concept is that auditors must report each verification result as a transaction in a blockchain, and the verification can always be timestamped just after the blockchain has recorded the transaction. In comparison to current systems, it has the drawback of continuous communication and computation overhead (Y.Zhang et al.,2019). Liet. al propose the CKshare network, which is built on private cloud and blockchain technology and promises to be a trustworthy mould redesign knowledge-sharing platform. Confidentiality, integrity, and availability are addressed by a symmetric encryption, hash-based connection and SHA256 encryption algorithms. However, there is no consensus process and smart contract included (Z. Li et al., 2019).

The parameters addressed in the above discussed survey papers has been analyzed and presented in Table 1. The table clearly presents the summary of the parameters addressed by the existing literature.

TABLE 1 Comparison analysis of existing blockchain survey papers

| **Paper** | **Applications** | **Domains** | **Consensus algorithm** | **Network architecture** | **Reward mechanisms** | **Scaling solutions** | **Services** | **Privacy issues/ Attacks** | **Challenges and Opportunities** |
|---|---|---|---|---|---|---|---|---|---|
| [10] | ✓ | ✓ | ✓ | ✓ | | | | ✓ | ✓ |
| [11] | ✓ | | | | | | | | |
| [12] | | | ✓ | | ✓ | | | | |
| [13] | | | | | | ✓ | | | |
| [3] | | | | | | | ✓ | | |
| [14] | | | | | | | | ✓ | |
| [7] | | | | | | | | ✓ | |
| [5] | ✓ | | ✓ | | | | | | ✓ |
| [15] | | | | | | | | ✓ | |
| [2] | ✓ | | ✓ | | | | | | ✓ |
| [9] | | | ✓ | | | | | | ✓ |
| [25] | ✓ | ✓ | ✓ | | | | | | |
| [26] | | | | | | | | ✓ | |

Baniata et. al explores and reviews published papers that combine blockchain and fog computing and categorized the papers based on their type, domain, publication year, blockchain function, consensus algorithm, and level where the blockchain has been configured for IoT, smart mobile devices applications, internet of vehicles applications, e-health applications, industrial internet of things applications (Baniata & Kertesz, 2020). Study of the interrelationship among blockchain and edge/fog computing, with a focus on solutions that are blockchain based is discussed. This discussion also covers privacy/security issues in the fog paradigm, as well as security requirements in fog enabled IoT framework using the blockchain (Tariq et. al, 2019). Wu et. al's proposed work customizes the blockchain for fog node clusters to reduce the necessary computing storage spaces and power consumption. This paper also proposes a novel approach to recover access control lists in blockchain-based fog node clusters. It also suggests the time aware computing set. allocation algorithm as a heuristic method for allocating computing power to calculate blocks' hash values which can reduce the overall processing time in obtaining the block's hash value. Each access control list update is treated as a single blockchain transaction. Each timeslot's access control list (ACL) hash value is used to synchronize all the ACLs stored in a single fog node cluster based on blockchain at the beginning of the following timeslot (D. Wu & Ansari, 2020).

The proposal of Blockchain and Fog-based Architecture Network (BFAN) would reduce fog node's average energy consumption, increase scalability, and achieve effective communications and computing and provide enhanced security in smart city applications. The data from smart sensors is treated as a blockchain transaction. The performance evaluation was carried out using the iFogSim simulator on the real data set (P. Singh et al., 2020). To address scalability problems, Kai Lei et. al proposes Groupchain, a public blockchain that can grow with a two-chain structure which is suitable for fog based IoT applications. It improves protection against attacks like double spend and selfish mining, as well as enhance transaction throughput and approval latency optimization (Lei et al., 2020). FogBus architecture, proposed in literature enables end to end IoT/fog/edge/cloud integration and provide platform-independent execution of IoT application. It helps developers create apps, aids users in running multiple apps at once and service providers in managing their services. Different FogBus configurations can adapt the computing environment to meet the needs of the situation. It is easily deployable, scalable, cost efficient, comparatively lightweight, and responsive (Tuli et al., 2019).

A distributed architecture of cloud based on blockchain and utilizing Software Defined Networking is proposed by Sharma et. al that allows controller fog nodes at the network's edge to fulfil the necessary design principles by providing less cost, reliable, and access to the IoT network's compute resources on demand. When compared to conventional IoT architecture, the effectiveness of the suggested system is enhanced by enhancing the IoT network's capacity to identify real-time attacks while lowering response times, increasing throughput, and minimizing end-to-end latency (Sharma et al.,2017). A suitable permission blockchain based on the fog system and a novel blockchain based fog architecture for the industrial IoT by splitting the proposed system structure into edge, fog, and cloud has been proposed. Transactions per second, active threads, Hyperledger based fog architecture's response time for elapsed time are evaluated in this proposal (Jang et al., 2019).

Bouachir et. al proposes an industrial cyber physical system that works on both blockchain and edge/fog approaches to address protection, quality of service, and data storage challenges while maintaining general data protection regulation (GDPR) complied privacy. Because of fog's decentralized computing resources, the distributed function of the blockchain can be better managed, resulting in increased scalability and system availability with a drawback of energy use, regulation, and standards (Bouachir et al., 2020). A lightweight hybrid federated learning architecture that uses blockchain tests the reliability of the distribution dataset when implemented for deep learning applications created for COVID patients in clinical trials. Smart contracts handle the edge trust, training, and authentication of federated nodes, and distribution dataset credibility. Every edge federated node conducts additive encryption and the blockchain aggregates the modified model parameters using multiplicative encryption. The differentially private internet of health things (IoHT) raw data is saved in the repository for authenticity and shared model training, and the training data hash position from the repository is stored in the blockchain (Rahman et al.,2020).

## 6. BLOCKCHAIN BASED FEDERATED LEARNING

Blockchains have the potential to decentralize the coordinating process for model generation in federated learning (Passerat-Palmbach et. al, 2020). Convergence is aided by blockchain-enabled federated learning, which allows for enhanced confirmations and participant selection (Drungilas et. al, 2021). FLchain is a blockchain based federated learning architecture proposed by Nguyen et. al that incorporates design issues and application cases (Nguyen et. al, 2021). BlockFL, a blockchain BAFFbased FL architecture proposed verifies and rewards local model improvements to exchange local model updates (H. Kim et. al, 2019). On a public blockchain network, Toyoda et. al propose a mechanism-design-oriented FL protocol, in which mechanism design is utilized to create a rule with a specific objective. It has the advantage of rewarding participants who have made significant contributions, which automatically discourages participants, deviating the protocol (Toyoda et. al, 2020). Blockchain based federated learning without aggregator (BAFFLE), a FL ecosystem is based on blockchain technology and is fundamentally decentralized, and uses smart contracts to manage round segmentation, aggregation of models, and updates in FL. After segmenting the world's parameter space into discrete parts, it uses a score and bid technique to speed up computation (Ramanan & Nakayama, 2020).

When compared to non-FL equivalents, Chain FL proposed by Korkmaz et. al uses the private blockchain to transfer the burden of storing the model to the network's nodes, produces promising outcomes with federated learning. To perform the federated learning update phase, miners run the smart contract code that was previously distributed on the network. Because the contract applies to all miners, system will continue to function even if one or more miners controlled by various parties fail. The smart contract installed in the private blockchain network allows any party to see both the most recent model and the past (Korkmaz et. al, 2020). A decentralized paradigm for cognitive computing powered by big data combines federated learning and blockchain to boost industry manufacturing 4.0 performance (Qu et. al, 2020). Two types of weights are proposed for selecting a subgroup of clients to update the global model using blockchain based federated learning. One type is by the weight based on accuracy of local learning of each client, and the other is by the weight based on each client's participation frequency. To compare the performance of the proposed system to that of existing schemes, important performance metrics such as learning speed and standard deviation were used (Y. J. Kim & Hong, 2019). Kim et. al employs a consensus mechanism in blockchain, on device machine learning without the need for centralized data for training or coordination. This encourages the federation of additional devices with a bigger number of training samples by delivering rewards proportionate to the training sample sizes. The BlockFL end to end latency model is developed by analyzing communication, computation, and PoW delays, and the subsequent latency is reduced by modifying the block production rate to accommodate the additional delay caused by the blockchain network (H. Kim et. al, 2019).

Incorporation of privacy and security preserving federated learning into a blockchain based architecture for secure data sharing between distributed parties has also been discussed. It incorporates federated learning within the permissioned blockchain consensus process, so that the computing power required for consensus can also be used for FL training (Lu et. al, 2019). For privacy-conscious and effective vehicular communication networking, on-vehicle machine learning model updates are communicated and verified in a distributed format utilizing an autonomous blockchain based federated learning architecture (Pokhrel & Choi, 2020). Aich et. al train the manufacturer's initial model using both the mobile edge server and the smart phone. Manufacturers designate consumers or organizations as second stage miners to compute the average model utilizing the models obtained from consumers. At the conclusion of the crowdsourcing project, one of the miners chosen as the temporary leader transfers the model to the blockchain to protect individuals' privacy, improve accuracy of test, provide differential privacy on the gathered characteristics, and introduce a novel normalization process (Aich et. al, 2022).

An asynchronous federated learning scheme proposed uses the edge data for learning and chooses the participating nodes carefully to reduce overall costs and integrate the learnt parameters into blockchain to increase the reliability of learned models and verify the properties of these parameters through two stage process (Lu et. al, 2020). FL-Block provides decentralized preservation of privacy while avoiding single points of failure with hybrid identity generation, full verification, access restriction, and data storage in off chain and retrieval (Qu et. al, 2020). Ma et. al studies about FL's shortcomings and further investigated blockchain assisted decentralized federated learning (BLADE-FL), a decentralized FL supported by blockchain, and demonstrated how effectively the suggested solution can handle possible problems, particularly the single point of failure issue, present in the conventional FL system (Kumar et. al, 2021). To protect the model parameters of B5G networks edge devices, proposal of combining blockchain based FL with wasserstein generative adversarial network enabled differential privacy. Blockchain makes it possible for decentralized FL to lower communication costs among the edge and cloud while resolving the problem of data falsification. It also offers a method for incentives to resolve the problem of data islands in beyond fifth generation (B5G)-driven edge computing (Wan et. al, 2022).

## 7. BLOCK CHAIN AND FEDERATED LEARNING FOR HEALTHCARE

Certain cognitive diseases can be revealed by some everyday behaviors. Wearable devices make it simple to access people's health records, with the help of rapid advancement in computing technology. The ability to learn from health data is constrained because of collection of technical ethical legal medical data privacy issues (Passerat-Palmbach et. al, 2019). The privacy and confidentiality of healthcare data must be secured from outside attackers since cybercriminals may be interested in such data for its confidential and sensitive information. A third-party vendor may utilize data analysis to categorize people who might be uninsurable because of their medical history or genetic abnormalities. Cybercriminals intending financial benefit from the theft of such data may sell the data to the vendor in exchange for a fee. To protect their patient's privacy, hospitals and other related organizations are unwilling to share data about their patients in real situations. Since Bitcoin's introduction, researchers have been working to expand blockchain's applications to non-financial use cases. Integrating blockchain technology into e-Health systems has the potential to increase service quality (Agbo et. al, 2019). Machine learning models are trained on a vast amount of consumer personal data to achieve great success in smart healthcare. Traditional machine learning models have privacy issues since original data is being sent for model training, and hence federated learning with more privacy can be incorporated with smart healthcare. In the smart health-care field, blockchain-based federated learning can be very effective.

Advantages of deploying a blockchain in healthcare sector approach is as follows:

- Without the use of a trusted mediator, an agreement may be achieved, preventing a performance issue and a single point of failure.
- Patients have rights to preserve their personal information.
- As blockchain data, medical history is accurate, secure, reliable, authentic, and distributed easily.
- Every member of the patient network can monitor the changes to the blockchain data and unauthorized modifications are easily detectable.

Blockchain technology has the potential to be disruptive, necessitating a complete rethink and substantial investment in the ecosystem (Esposito et. al, 2018). Using blockchain technology, a secure cloud-assisted eHealth framework is proposed by Cao et. al to defend outsourced Electronic Health Records (EHRs) from unauthorized alteration. According to the suggested method, only authenticated participants are permitted to outsource EHRs, and each such activity generates a transaction on a public, blockchain (Cao et. al, 2019). In terms of power consumption, latency, network use, and load balancing, a smart remote healthcare system with and without blockchain has been compared (Ejaz et. al, 2021). Islam et. al proposea system for activity tracking and identification based on a multi class cooperative procedure of categorization which is deployable either online offline to improve the activity classification's accuracy in aid of a blockchain architecture based on fog or cloud computing. A method that uses the frame-based prominent characteristics of videos of various human behaviors, which are then processed using a support vector machine that is built on the idea of error correction output codesfor efficiency and accuracy (Islam et. al, 2019).

Chen et. al propose FedHealth, system that use FL to aggregate data and then uses transfer learning to create models that are personalized. Transfer learning is used to adjust the model to the user because there is a high distribution divergence between data of server and user where all the parameter exchange processes use homomorphic encryption to ensure that no user data is leaked. The design of a new secure aggregation protocol includes a secure hardware component and an Ethereum-native encryption toolkit (Chen et. al, 2020). A system proposed in literature gathers a modest quantity of data from different hospitals and train a global model based on blockchain using deep learning enabled federated learning (Kumar et. al, 2021). A secure architecture for smart healthcare that is supported by blockchain and federated learning has been proposed which protects privacy by using blockchain-based IoT cloud platforms (S Singh et. al, 2022). Recent FL designs for smart healthcare include resource-aware FL, secure and privacy-aware FL, incentive FL, and personalised FL. Presentation of a state-of-the-art review of the new FL applications in vital areas of healthcare, such as COVID-19 detection, medical imaging, and remote health monitoring has been discussed (Nguyen et. al, 2022)

Principal advantages of federated learning in healthcare sector approach are as follows:

- Individual hospitals can benefit from the vast datasets of numerous hospitals using federated learning instead of centralizing the data in one location.
- Auto-scaling is enabled by the federated learning healthcare systems at absolutely no added expense.
- Helps to solve the challenge of healthcare data regulation and privacy by collectively training algorithms without disclosing the data.
- It enables precision medicine by allowing models to make balanced decisions that reflect aperson's individual physiology while considering governance and security/privacy concerns.

A classification of the articles that were reviewed for the survey is presented in Figure 8. Comparison of parameters in survey papers of edge/fog/cloud/IoT healthcare applications analyzed in section 5, 6 and 7 are described in Table 2.

**TABLE 2 Comparison of parameters in Edge/Fog/Cloud/IoT systems**

| **Paper** | **Applications** | **Consensus Algorithm** | **Transaction** | **Type** | **Reward mechanisms** | **Smart contracts** | **Advantages** |
|---|---|---|---|---|---|---|---|
| **[27]** | Fog computing applications | PoW | Each Access control list entry | Public | No | No | Manage computing power |
| **[28]** | Smart city | Not mentioned | The information obtained by smart sensors | Public | No | No | Improved security features are ensured while the latency and energy are reduced. |
| **[29]** | Fog Computing ofIoT Services | PoW | Data from IoT nodes | Public | Yes | Yes | Optimization on transaction throughput and confirmation latency |
| **[30]** | Framework for executing and deploying apps for various IoT-enabled systems | PoW | Data from different IoT enabled systems | Not mentioned | No | No | Enables the deployment, management, and monitoring of IoT applications and resources |
| **[31]** | Big data producing IoT applications | Proof of Service | Task completion, data management, and server provision | Private | Yes | Yes | Enables Economical Highly effective computing |
| **[32]** | Industrial IoT | Not mentioned | Data from the recently registered edge device is queried and exchanged for resources. | Public | No | Yes | Guarantee fast performance |
| **[33]** | Cyber physicalsystems | Not mentioned | An IIoT device requests a transaction, after which the device's metadata are added to a block. | Public or Private | No | No | Increased scalability and system availability |
| **[36]** | Healthcare | PoS and PoW | Electronic health records | Not mentioned | No | No | Guarantees confidentiality, correctness, integrity of EHRs |

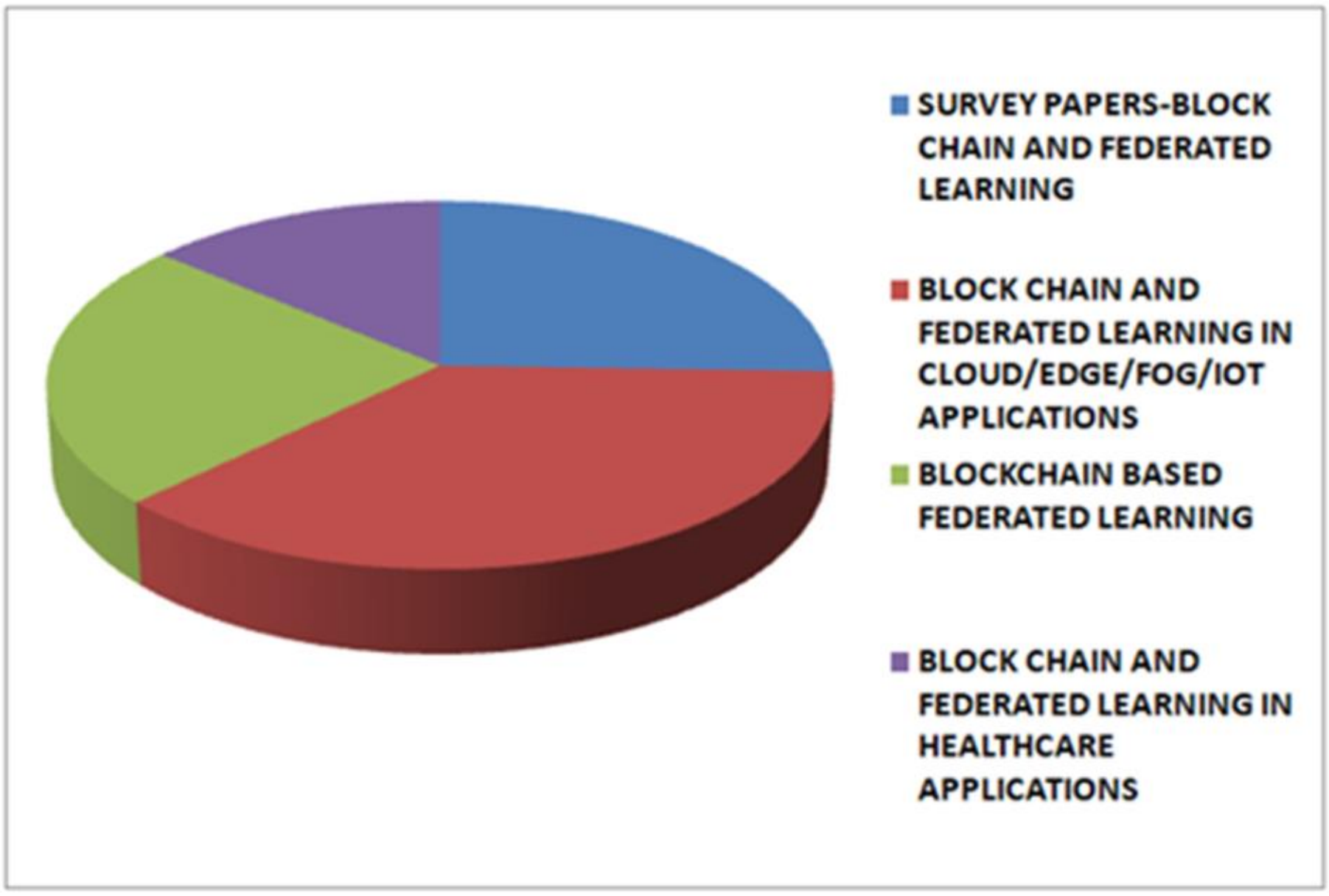

**FIGURE 8 A classification of the articles that were reviewed for the survey**

## 8. CONCLUSION

This paper explores the significance of federated learning with blockchain integration for IoT healthcare solutions in Edge/Fog/Cloud computing environments. Our survey is divided into different sections: architecture, benefits of federated learning with blockchain integration, blockchain based federated learning, blockchain and federated learning application areas, blockchain and federated learning in healthcare applications touching upon numerous aspects of the relationship between blockchain/federated learning and edge/fog and IoT systems. The extensive and thorough review of the blockchain/FL architecture, as well as its features, are discussed in the first section. In the subsequent sections, the discussion turns to focus more on blockchain/FL in edge/fog/cloud computing in IoT and healthcare applications. For future smart implementations, the research gaps in blockchain-based federated learning have been identified.

To the best of our knowledge, there is currently no work that offers a detailed and systematic evaluation of blockchain, and federated learnings use in cloud/edge/fog IoT networks and applications, despite the fact that blockchain and federated learning have been extensively discussed in the literature.